\documentclass[journal]{IEEEtran}
\usepackage{ifpdf}

\usepackage{cite}

\ifCLASSINFOpdf
   \usepackage[pdftex]{graphicx}
\else
\fi
\usepackage[cmex10]{amsmath}
\usepackage{amsmath}
\usepackage{amssymb}

\usepackage{algorithmic}
\usepackage{algorithm}

\usepackage{array}

\usepackage{mdwmath}
\begin{document}
%
\title{MIMO UWB Radar System with Compressive Sensing}


\author{Xia Li,~\IEEEmembership{Student Member,~IEEE,}      
				Zhen Hu,
        Robert C. Qiu,~\IEEEmembership{Fellow,~IEEE,}
				
\thanks{X. Li, Z. Hu, and R. C. Qiu are with the Department of Electrical and Computer Engineering, Center for Manufacturing Research, Tennessee Technological University, Cookeville, TN, 38505, e-mail: xli43@students.tntech.edu; zhu@tntech.edu; rqiu@tntech.edu.}
}

\maketitle

\begin{abstract}


A multiple input multiple output ultra-wideband cognitive radar based on compressive sensing is presented in this letter. For traditional UWB radar, high sampling rate analog to digital converter at the receiver is required to meet Shannon theorem, which increases hardware complexity. In order to bypass the bottleneck of ADC or further increase the radar bandwidth using the latest wideband ADC, we propose to exploit CS for signal reconstruction at the receiver of UWB radar for the sparse targets in the surveillance area. Besides, the function of narrowband interference cancellation is integrated into the proposed MIMO UWB radar. The field demonstration proves the feasibility and reliability of the proposed algorithm.

\end{abstract}


%
\IEEEpeerreviewmaketitle

\section{Introduction}



In the radar system, the unprecedented radio bandwidth provides advantages such as the high precision of range estimation. However, the extremely high sampling rate of analog to digital converter (ADC) becomes a major challenge in the ultra-wideband (UWB) radar systems. According to Nyquist sampling theorem, the sampling rate should be at least twice the bandwidth of the signal. For better system performance, oversampling is often required and results even higher sampling rate. Compressed sensing (CS) gives an opportunity to overcome this challenge. CS~\cite{candes2006robust} theory specifies a new signal acquisition approach, in a high probability, allowing the acquisition of signals at a much lower data rate than the Nyquist sampling rate.

Some previous work of our group~\cite{zhang2009compressed} applied CS to the communication application. In the communication system, the received waveforms are synchronized by some mechanism like preamble. However, in the MIMO radar application, the received signal cannot be considered as synchronized since the arrival time of signal depends on the over the air delay, which is determined by the target distance to the radar. The unknown delay leads to the randomness of the measurement matrix construction.  A cyclic prefix (CP) is used in waveform design to make sure the measurement matrix is constant and unrelated to the time delay. Therefore, each receiver antenna can construct the basis matrix locally.  Based on the analysis of the system, the architecture of the testbed is proposed. This testbed is based on our previous results of the UWB time reversal experiment~\cite{chen2012towards} and cognitive UWB radar experiment~\cite{li2012waveform, li2014demonstration, li2012experimental}. Both of them are on the full sampling rate. However, the new system satisfies the specialty of the compressed sensing structure.  The sampling rate at receiver can be much lower than the Nyquist rate used in the two previous systems.


The experiment results based on measured data are provided to demonstrate the effectiveness of the algorithm. The performance curves of the algorithm under different NBI and SNR conditions are provided in simulation.

\section{Background and Problem Formulation}
\label{System_Background}

\def\conv{\mathbin{\hbox{$\,*$\kern-1.5ex$\odot$}}}

\begin{figure}[!t]
	\centering
	\includegraphics[width=2.8in]{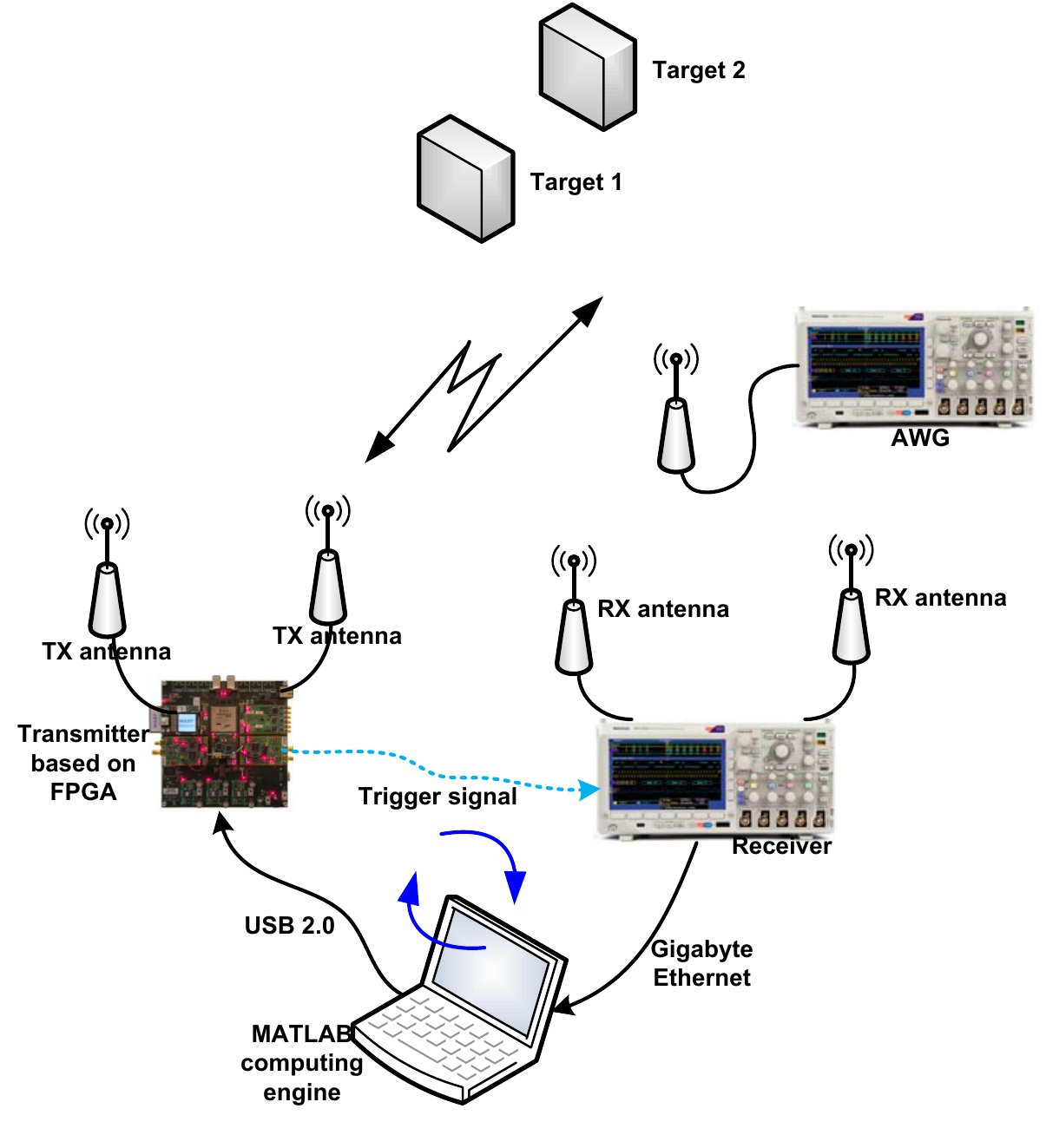}
	\caption{Block diagram of the system architecture}
	\label{DeviceSetup}
\end{figure}

Since the transmitter and receiver of the our system are synchronized by a trigger signal as shown in Fig.~\ref{DeviceSetup}, ToAs between transmitter antennas and receiver antennas can be estimated from the recovered channel information, which are used to estimate the target positions\cite{li2012waveform}.

In traditional channel estimation theory, the channel vector ${\mathbf{h}} \in {\mathbb{C}^n}$ can be estimated from the observations of the linear convolution, ${\mathbf{Xh}} = {\mathbf{y}} \in {\mathbb{C}^n}$., where the observation matrix ${\mathbf{X}} \in {\mathbb{C}^{n \times n}}$ is a partial Toeplitz matrix.

By using CS theory, a sparse, high-dimensional vector ${\mathbf{h}} \in {\mathbb{C}^n}$ can be accurately recovered from a small number of linear observations, ${{\mathbf{R}}_\Omega }{\mathbf{Xh}} = {\mathbf{\bar y}} \in {\mathbb{C}^m}$. Consider an arbitrary index set $\Omega  \subset \left\{ {0,1, \ldots ,n - 1} \right\}$ whose cardinality $\left| \Omega  \right| = m$. We define the operator ${{\mathbf{R}}_\Omega } \in {\mathbb{C}^{m \times n}}$ that restricts a vector to the entries listed in $\Omega$. In our testbed, the ${\mathbf{\bar y}}$ can be viewed as the sub-sampling of ${\mathbf{y}}$. A delay in ${\mathbf{\bar y}}$ means the change of ${{\mathbf{R}}_\Omega }$, which means the measurement matrix ${{\mathbf{R}}_\Omega }{\mathbf{X}}$ is different. Since the delay is unknown to the receiver, channel information cannot be recovered at receiver locally.

A cyclic extension of time length ${T_G}$ is used to change the linear convolution to circulant convolution. If the channel delay is within ${T_G}$, the system can be described in matrix as ${\mathbf{y}} = {\mathbf{Xh}} + {\mathbf{n}}$, where $\mathbf{X}$ is a circulant matrix as
\begin{equation}
\label{Toeplitz_matrix}
{\mathbf{X}} = \left[ {\begin{array}{*{20}{c}}
  {{x_{n - 1}}}&{{x_{n - 2}}}& \cdots &{{x_1}}&{{x_0}} \\ 
  {{x_0}}&{{x_{n - 1}}}& \ldots &{{x_2}}&{{x_1}} \\ 
   \vdots & \vdots & \vdots & \vdots & \vdots  \\ 
  {{x_{n - 2}}}&{{x_{n - 3}}}& \cdots &{{x_0}}&{{x_{n - 1}}}
\end{array}} \right].
\end{equation}

The delay in dictionary ${n_{{\rm{d}}}}$ can be represented by the channel latency. This is proved by (\ref{CP}). It holds when the delay ${n_{{\rm{d}}}}$ is smaller than the length of CP.
\begin{eqnarray}
\label{CP}
{y_c}\left[n - {n_{{\rm{d}}}}\right] &=& \sum\limits_{m = 0}^{N - 1} {h\left[m\right]x\left[{{\left\langle {n - {n_{{\rm{d}}}} - m} \right\rangle }_N}\right]} \nonumber\\
 &=& \sum\limits_{m = 0}^{N - 1} {h\left[m\right]x\left[{{\left\langle {n - ({n_{{\rm{d}}}} + m)} \right\rangle }_N}\right]} \nonumber\\
 &=& \sum\limits_{m' = 0}^{N - 1} {h\left[m' - {n_{{\rm{d}}}}\right]x\left[{{\left\langle {n - m'} \right\rangle }_N}\right]} \nonumber\\
 &=& \sum\limits_{m = 0}^{N - 1} {h\left[m - {n_{{\rm{d}}}}\right]x\left[{{\left\langle {n - m} \right\rangle }_N}\right]} 
\end{eqnarray}

The vector ${\mathbf{h}}$ is the sampled signal of the channel impulse response $h\left( t \right)$. $h\left( t \right)$ can be treated as a time limited pulse train of the form $h\left( t \right) = \sum\limits_m {{\alpha _m}\delta \left( {t - {\tau _m}{T_s}} \right)}$, where the amplitudes ${\alpha _m}$ of channel taps are complex valued.
The samples of ${\mathbf{h}}$ are modeled as~\cite{van1995channel}
\begin{equation}
\label{channel_model}
{h_k} = \frac{1}{{\sqrt n }}\sum\limits_m {{\alpha _m}{e^{ - j\frac{\pi }{N}\left( {k + \left( {N - 1} \right){\tau _m}} \right)}}\frac{{\sin \left( {\pi {\tau _m}} \right)}}{{\sin \left( {\frac{\pi }{N}\left( {{\tau _m} - k} \right)} \right)}}}.
\end{equation}
If the delay ${{\tau _m}}$ is an integer, then all the energy from ${{\alpha _m}}$ is mapped to tap ${h_{{\tau _m}}}$. However, if ${{\tau _m}}$ is not an integer, its energy will leak to all taps of ${{h_k}}$. However, most of the energy is kept in the neighborhood of the original pulse locations. When the target number is small, the number of wireless paths is also small. In this case, ${\mathbf{h}}$ can be considered as a sparse signal.

For a $2 \times 2$ MIMO system shown in Fig.~\ref{DeviceSetup}, the system equation can be described as
\begin{align}
\label{MIMO_matrix}
  \left[ {\begin{array}{*{20}{c}}
  {{{\mathbf{y}}_1}}&{{{\mathbf{y}}_2}} 
\end{array}} \right] &= \left[ {\begin{array}{*{20}{c}}
  {{{\mathbf{X}}_1}}&{{{\mathbf{X}}_2}} 
\end{array}} \right]\left[ {\begin{array}{*{20}{c}}
  {\underbrace {\left[ {\begin{array}{*{20}{c}}
  {{{\mathbf{h}}_{11}}} \\ 
  {{{\mathbf{h}}_{12}}} 
\end{array}} \right]}_{{{\mathbf{h}}_1}}}&{\underbrace {\left[ {\begin{array}{*{20}{c}}
  {{{\mathbf{h}}_{21}}} \\ 
  {{{\mathbf{h}}_{22}}} 
\end{array}} \right]}_{{{\mathbf{h}}_2}}} 
\end{array}} \right] \nonumber\\ 
   & \quad + \left[ {\begin{array}{*{20}{c}}
  {{{\mathbf{n}}_1}}&{{{\mathbf{n}}_2}} 
\end{array}} \right],
\end{align}
where ${{{\mathbf{y}}_1}}$ and ${{{\mathbf{y}}_2}}$ are the received waveforms from receiver antennas 1 and 2. For the transmitted waveforms ${{{\mathbf{x}}_1}}$ and ${{{\mathbf{x}}_2}}$, ${{{\mathbf{X}}_1}}$ and ${{{\mathbf{X}}_2}}$ are the respective Toeplitz matrices. ${{{\mathbf{n}}_1}}$ and ${{{\mathbf{n}}_2}}$ are the i.i.d. complex Gaussian variables for the two receiver antennas. ${{\mathbf{h}}_{ji}}$ is the impulse response of the channel between transmitter antenna $i$ to receiver antenna $j$.

For the CS-based radar system, the sub-sampled received waveform can be considered as the result of down-sampling of the received waveform with full sampling rate.
\begin{align}
\label{MIMO_matrix_CS}
  \left[ {\begin{array}{*{20}{c}}
  {{{{\mathbf{\bar y}}}_1}}&{{{{\mathbf{\bar y}}}_2}} 
\end{array}} \right] &= \left[ {\begin{array}{*{20}{c}}
  {\overline {{{\mathbf{X}}}}_1 }&{\overline {{{\mathbf{X}}}}_2 } 
\end{array}} \right]\left[ {\begin{array}{*{20}{c}}
  {\underbrace {\left[ {\begin{array}{*{20}{c}}
  {{{\mathbf{h}}_{11}}} \\ 
  {{{\mathbf{h}}_{12}}} 
\end{array}} \right]}_{{{\mathbf{h}}_1}}}&{\underbrace {\left[ {\begin{array}{*{20}{c}}
  {{{\mathbf{h}}_{21}}} \\ 
  {{{\mathbf{h}}_{22}}} 
\end{array}} \right]}_{{{\mathbf{h}}_2}}} 
\end{array}} \right] \nonumber\\ 
   &\quad + \left[ {\begin{array}{*{20}{c}}
  {{{{\mathbf{\bar n}}}_1}}&{{{{\mathbf{\bar n}}}_2}} 
\end{array}} \right]  
\end{align}


The corresponding partial circulant matrix is defined as ${\mathbf{\bar X}} = {{\mathbf{R}}_\Omega }{\mathbf{X}}$. The partial random circulant matrix is generated by a Rademacher sequence. It is proved in~\cite{rauhut2012restricted}, the $s$th restricted isometry constant ${\delta _s}$ has 
\begin{equation}
\label{RPI}
\mathbb{E}\left[ {{\delta _s}} \right] \leqslant {C_1}\max \left\{ {\frac{{{s^{\frac{3}{2}}}}}{m}{{\log }^{\frac{3}{2}}}n,\sqrt {\frac{s}{m}} \log s\log n} \right\},
\end{equation}
where ${C_1} > 0$ is a universal constant. 

~(\ref{RPI}) means that for a given $\delta  \in \left( {0,1} \right)$, we have $\mathbb{E}\left[ {{\delta _s}} \right] \leqslant \delta $ provided by
\begin{equation}
m \geqslant {C_2}\max \left\{ {{\delta ^{ - 1}}{s^{\frac{3}{2}}}{{\log }^{\frac{3}{2}}}n,{\delta ^{ - 2}}s{{\log }^2}n{{\log }^2}s} \right\},
\end{equation}
where ${C_2} > 0$ is another universal constant.

For $0 \leqslant \lambda  \leqslant 1$ and ${C_3} > 0$, we can have 
\begin{equation}
\mathbb{P}\left( {{\delta _s} \geqslant \mathbb{E}\left[ {{\delta _s}} \right] + \lambda } \right) \leqslant {e^{\frac{{ - {\lambda ^2}}}{{{\sigma ^2}}}}},
\end{equation}
where ${\sigma ^2} = {C_3}\frac{s}{m}{\log ^2}s{\log ^2}n$.

In radar systems, the number of multipath is equivalent to the number of targets. The channel information is sparse to the most cases. The channel information can be recovered by using the following convex optimization program 
\begin{equation}
\label{system_convex}
\begin{gathered}
  \min \left( {{{\left\| {{{\mathbf{h}}_1}} \right\|}_1} + {{\left\| {{{\mathbf{h}}_2}} \right\|}_1}} \right)\; \hfill \\
  {\text{subject to }}\left\{ {\begin{array}{*{20}{c}}
  {{{\left\| {\overline{\mathbf{ X}}{{\mathbf{h}}_1} - {\mathbf{\bar y}_1}} \right\|}_2} \leqslant {\varepsilon _1}} \\ 
  {{{\left\| {\overline{\mathbf{ X}}{{\mathbf{h}}_2} - {\mathbf{\bar y}_2}} \right\|}_2} \leqslant {\varepsilon _2}} 
\end{array}} \right. \hfill \\ 
\end{gathered},
\end{equation}
where ${\left\|  \cdot  \right\|_1}$ is the $\ell_1$ norm and ${\left\|  \cdot  \right\|_2}$ is the $\ell_2$ norm. ${\varepsilon _j}$ is chosen such that it bounds the amount of noise in the measurement from $j$th receiver antenna. The convex optimization problem~(\ref{system_convex}) is a second-order cone problem and can be solved efficiently by utilizing the Basis Pursuit Denoising (BPDN) algorithm~\cite{chen2001atomic}. 
 

\section{NBI cancellation using CS theory}
\label{NBI_cancel}

The system is modified to achieve the robustness under NBI. In~(\ref{MIMO_matrix}), NBI signal is not taken into account in algorithm,  NBI is treated as noise. However, the sparsity of the NBI can be used to recover its information from sub-sampled received waveform.

\begin{align}
\label{MIMO_matrix_NBI}
  \left[ {\begin{array}{*{20}{c}}
  {{{{\mathbf{\bar y}}}_1}}&{{{{\mathbf{\bar y}}}_2}} 
\end{array}} \right] &= {\overline {{{\mathbf{X}}}} }\left[ {\begin{array}{*{20}{c}}
  {{{\mathbf{h}}_1}}&{{{\mathbf{h}}_2}} 
\end{array}} \right] + \left[ {\begin{array}{*{20}{c}}
  {{{{\mathbf{\bar j}}}_1}}&{{{{\mathbf{\bar j}}}_2}} 
\end{array}} \right] \nonumber\\ 
   &\quad + \left[ {\begin{array}{*{20}{c}}
  {{{{\mathbf{\bar n}}}_1}}&{{{{\mathbf{\bar n}}}_2}} 
\end{array}} \right] \nonumber\\ 
   &= {\overline {{{\mathbf{X}}}} }\left[ {\begin{array}{*{20}{c}}
  {{{\mathbf{h}}_1}}&{{{\mathbf{h}}_2}} 
\end{array}} \right] + \left[ {\begin{array}{*{20}{c}}
  {\overline {{{\mathbf{F}}^H}} {{\mathbf{j}}_{\text{f}}}_1}&{\overline {{{\mathbf{F}}^H}} {{\mathbf{j}}_{\text{f}}}_2} 
\end{array}} \right] \nonumber\\ 
   &\quad+ \left[ {\begin{array}{*{20}{c}}
  {{{{\mathbf{\bar n}}}_1}}&{{{{\mathbf{\bar n}}}_2}} 
\end{array}} \right] 
\end{align}
where $\overline {{{\bf{F}}^H}}$ is a randomly chosen rows of the discrete Fourier matrix and can be written as $\overline {{{\bf{F}}^H}}  = {{\bf{R}}_\Omega }{{\bf{F}}^H}$. ${{{{\bf{ j}}}_j}}$ is a vector of the NBI signal received at antenna $j$. ${{\bf{j}}_{\rm{f}}}_j$ is a vector of the NBI in frequency domain. A recent research~\cite{candes2011compressed} has proved that this measurement matrix has small restricted isometry constants with very high probability.

For the receiver antenna $j$, we can have
\begin{align}
\label{MIMO_matrix_NBI_convex}
  {{{\mathbf{\bar y}}}_j} &= \left[ {\begin{array}{*{20}{c}}
  {\overline {{{\mathbf{X}}}}_1 }&{\overline {{{\mathbf{X}}}}_2} 
\end{array}} \right]\left[ {\begin{array}{*{20}{c}}
  {{{\mathbf{h}}_{j1}}} \\ 
  {{{\mathbf{h}}_{j2}}} 
\end{array}} \right] + {{\mathbf{F}}^{ - 1}}{{\mathbf{j}}_{\text{f}}}_j + {{\mathbf{\bar n}}_j} \nonumber\\ 
   &= \underbrace {\left[ {\begin{array}{*{20}{c}}
  {\begin{array}{*{20}{c}}
  {\overline{{{\mathbf{X}}}}_1 }&{\overline{{{\mathbf{X}}}}_2}
\end{array}}&{\overline {{{\mathbf{F}}^H}} } 
\end{array}} \right]}_{{\mathbf{\bar A}}}\underbrace {\left[ {\begin{array}{*{20}{c}}
  {{{\mathbf{h}}_{j1}}} \\ 
  {{{\mathbf{h}}_{j2}}} \\ 
  {{{\mathbf{j}}_{\text{f}}}_j} 
\end{array}} \right]}_{{{\mathbf{s}}_j}} + {{\mathbf{\bar n}}_j}. 
\end{align}

Because both the channel information and the NBI interference information are sparse, the interference signal and channel information can be recovered  simultaneously by using the following convex optimization program
\begin{equation}
\label{MIMO_matrix_NBI_convex_opti}
\begin{gathered}
  \min \left( {{{\left\| {{{\mathbf{s}}_1}} \right\|}_1} + {{\left\| {{{\mathbf{s}}_2}} \right\|}_1}} \right)\;{\kern 1pt}  \hfill \\
  {\text{subject to }}\left\{ {\begin{array}{*{20}{c}}
  {{{\left\| {\overline{\mathbf{ A}}{{\mathbf{s}}_1} - {{{\mathbf{\bar y}}}_1}} \right\|}_2} \leqslant {\varepsilon' _1}} \\ 
  {{{\left\| {\overline{\mathbf{ A}}{{\mathbf{s}}_2} - {{{\mathbf{\bar y}}}_2}} \right\|}_2} \leqslant {\varepsilon' _2}} 
\end{array}} \right. \hfill \\ 
\end{gathered} ,
\end{equation}
where ${{\mathbf{s}}_1}$ contains both the channel information and interference position in frequency domain. ${{\varepsilon '_j}}$ bounds the energy of ${{\mathbf{\bar n}}_j}$.

NBI signal is always sparse in frequency domain due to its narrow band feature. The number of taps in channel information equals the number of targets in the surveillance area and suffers the influence of the environment like clutters. It is more likely  NBI has better sparsity structure than the channel information. The precision of the recovery algorithm relies on the signal sparsity. The NBI signal recovery should be more accurate. Thus, the estimated interference signal ${{{\mathbf{\tilde j}}}_{\text{f}}}$ can be used to cancel the interference in the received waveform. By subtracting the interference signal ${{\mathbf{j}}_{\text{f}}}$ using the estimated interference signal ${{{\mathbf{\tilde j}}}_{\text{f}}}$, the remaining signal ${{\mathbf{\bar w}}}$ is obtained.

\begin{align}
\label{MIMO_matrix_NBI_convex_opti_cancel}
  {{{\mathbf{\bar y'}}}_j} &= \left[ {\begin{array}{*{20}{c}}
  {\overline {{{\mathbf{X}}}}_1 }&{\overline {{{\mathbf{X}}}}_2 } 
\end{array}} \right]\left[ {\begin{array}{*{20}{c}}
  {{{\mathbf{h}}_{j1}}} \\ 
  {{{\mathbf{h}}_{j2}}} 
\end{array}} \right] + \overline {{{\mathbf{F}}^H}} {{\mathbf{j}}_{\text{f}}}_j + {{{\mathbf{\bar n}}}_j} - \overline {{{\mathbf{F}}^H}} {\tilde {\bf{j}}_{{\rm{f}}j}} \nonumber\\ 
   &= \underbrace {\left[ {\begin{array}{*{20}{c}}
  {\overline {{{\mathbf{X}}}}_1 }&{\overline {{{\mathbf{X}}}}_2 } 
\end{array}} \right]}_{{\mathbf{\bar X}}}\underbrace {\left[ {\begin{array}{*{20}{c}}
  {{{\mathbf{h}}_{j1}}} \\ 
  {{{\mathbf{h}}_{j2}}} 
\end{array}} \right]}_{{{\mathbf{h}}_j}} + {{{\mathbf{\bar w}}}_j} + {{{\mathbf{\bar n}}}_j}
\end{align}

The channel information can be recovered by using the following convex optimization program
\begin{equation}
\label{MIMO_matrix_NBI_convex_opti_cancel_convx}
\begin{gathered}
  \min \left( {{{\left\| {{{\mathbf{h}}_1}} \right\|}_1} + {{\left\| {{{\mathbf{h}}_2}} \right\|}_1}} \right)\;{\kern 1pt}  \hfill \\
  {\text{subject to }}\left\{ {\begin{array}{*{20}{c}}
  {{{\left\| {\overline{\mathbf{ X}}{{\mathbf{h}}_1} - {{{\mathbf{\bar y'}}}_1}} \right\|}_2} \leqslant {\varepsilon'' _1}} \\ 
  {{{\left\| {\overline{\mathbf{ X}}{{\mathbf{h}}_2} - {{{\mathbf{\bar y'}}}_2}} \right\|}_2} \leqslant {\varepsilon'' _2}} 
\end{array}} \right. \hfill \\ 
\end{gathered} ,
\end{equation}
where ${{\varepsilon ''_j}}$ bounds the energy of noise signal ${{{\mathbf{\bar w}}}_j} + {{{\mathbf{\bar n}}}_j}$.

\section{CS-based MIMO radar system}
\label{radar_system}



\begin{figure}[!t]
	\centering
	\includegraphics[width=3.2in]{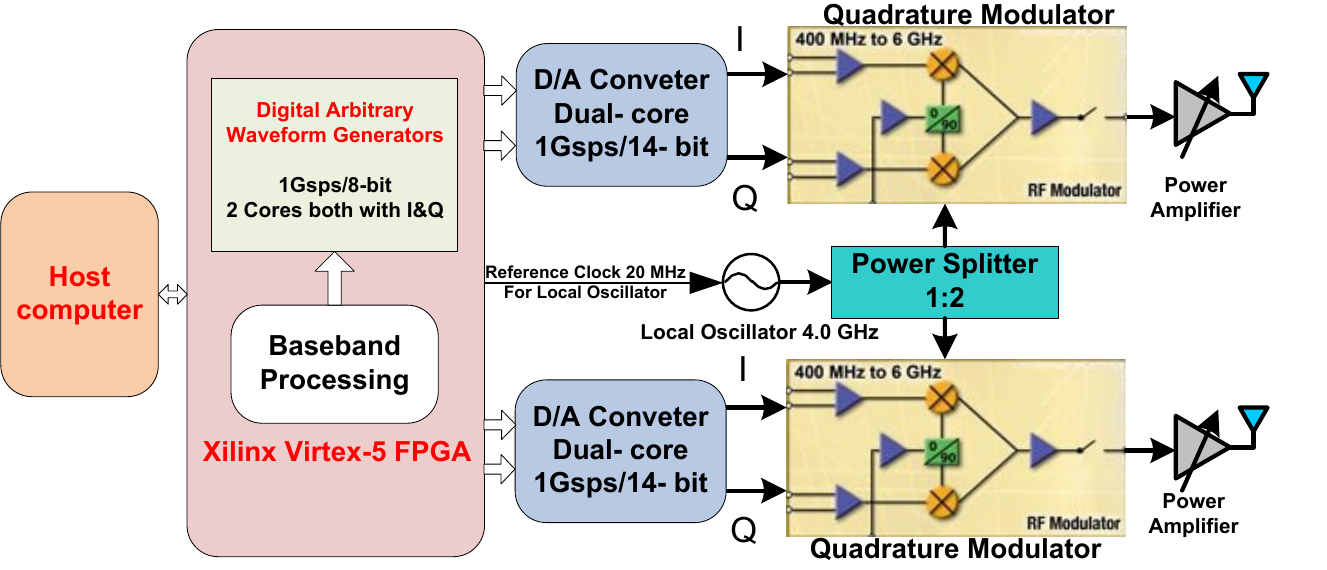}
	\caption{Block diagram of the system architecture}
	\label{transmitter}
\end{figure}

Based on the analysis on the previous section, a CS-based cognitive MIMO radar system is proposed. The transmitter architecture is the same as our previous work~\cite{chen2012towards}, as shown in Fig.~\ref{transmitter}. In the experiment, the carrier frequency is set to 4 GHz. The bandwidth of the radar waveform is 500 MHz, making the system an ultra-wideband system.

\begin{figure}[!t]
	\centering
	\includegraphics[width=3.2in]{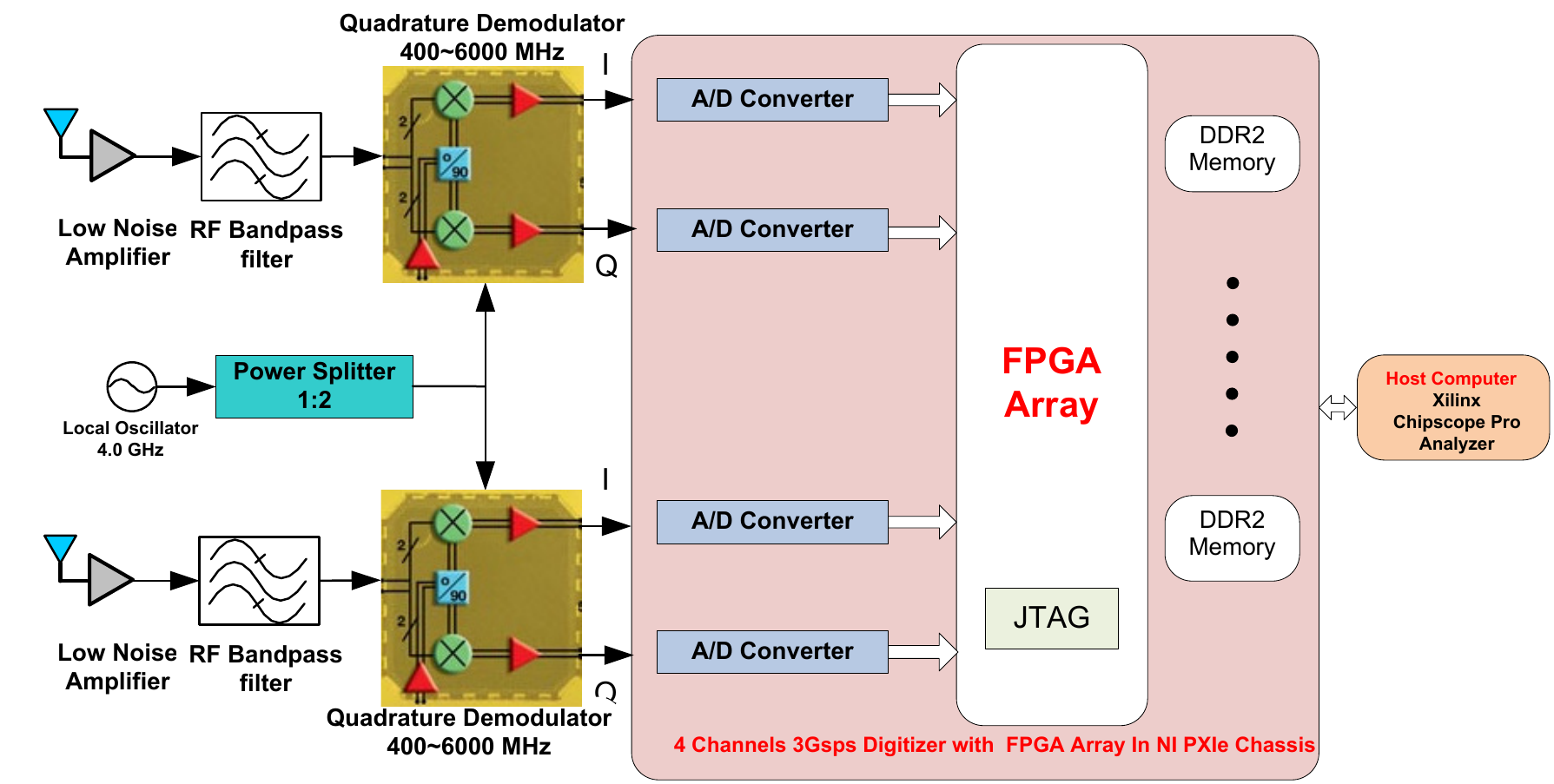}
	\caption{Block diagram of the system architecture}
	\label{receiver}
\end{figure}

To recover the original signal, the sampling rate of the radar received waveform must be higher than 500 MHz in traditional system. Due to the CS algorithm, the sampling rate of the received signal can be reduced to much lower than the Nyquist rate. As an example, we set the receiver ADC sampling rate $\frac{1}{3}$ of the full signal bandwidth, which is 166.66 Msps, as shown in Fig.~\ref{receiver}. The reduced sampling rate in the receiver can greatly reduce the cost and power consumption of the receiver.  The analog RF down-converter is the same as the traditional system. A bandpass filter is used to remove noise outside of signal bandwidth.

\section{Experiment Results and Performance Analysis}
\label{performance_analysis}

\begin{figure}[!t]
	\centering
	\includegraphics[width=2.8in]{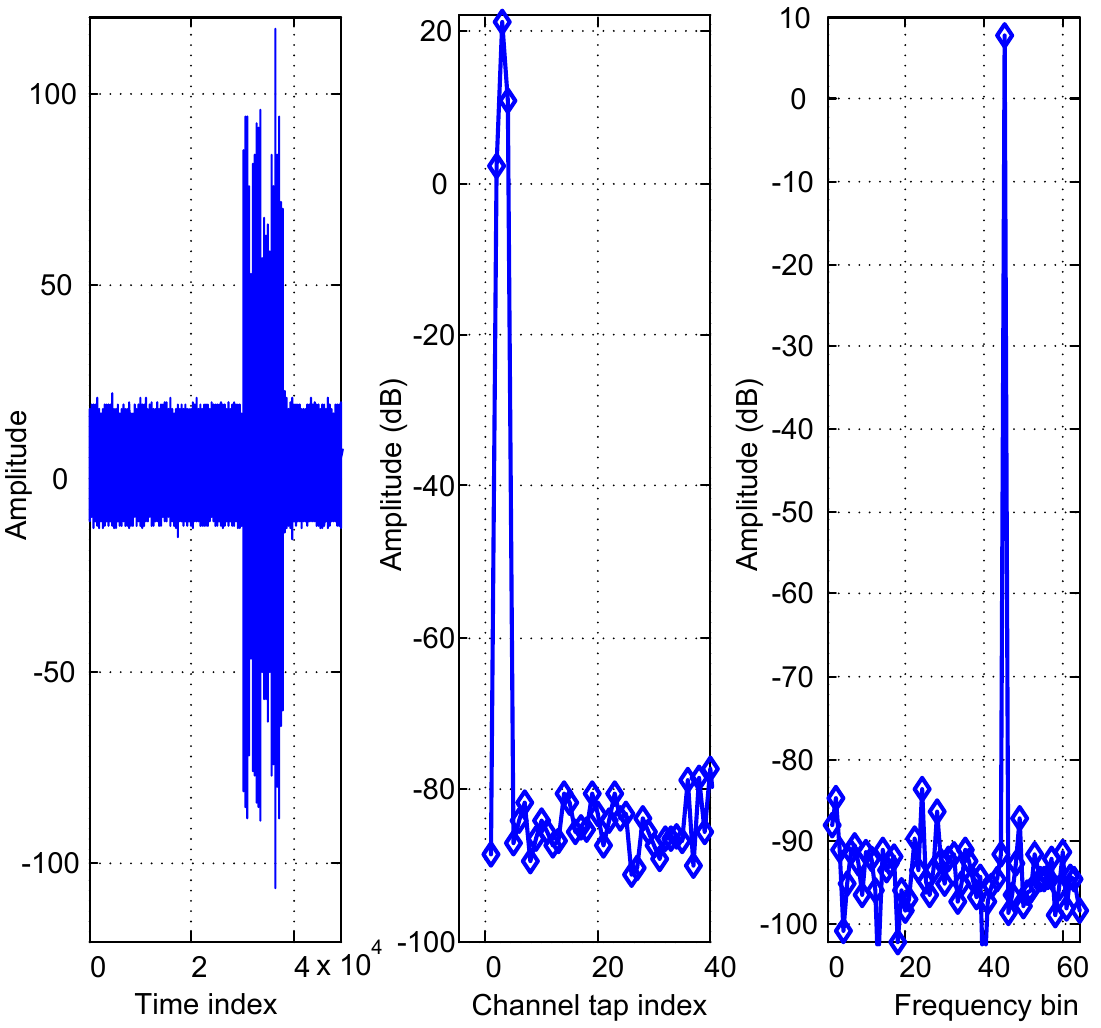}
	\caption{Experiment data: (a) Received sub-sampled baseband signal; (b) Recovered channel information; (c) Recovered interference information}
	\label{recovered}
\end{figure}

With one target in surveillance area, the experiment result is shown in Fig.~\ref{recovered}.  The starting time of recording received signal is the starting time of waveform transmitting. The position of radar waveform in the received signal can be detected by using the correlation relationship of CP. The channel information should have only one path in this situation. However, in real system, the channel delay is not an integer in model~(\ref{channel_model}). Fig.~\ref{recovered} (b) is the result by solving~(\ref{MIMO_matrix_NBI_convex_opti_cancel_convx}). Most energy is mapped to tap 3 while the neighborhood of the tap holds the residue. Since sparsity is considered, the values of all other taps are close to zero. Fig.~\ref{recovered} (c) Interference information is obtained by solving~(\ref{MIMO_matrix_NBI_convex}). All of the energy is mapped to tap 45.

\begin{figure}[!t]
	\centering
	\includegraphics[width=2.8in]{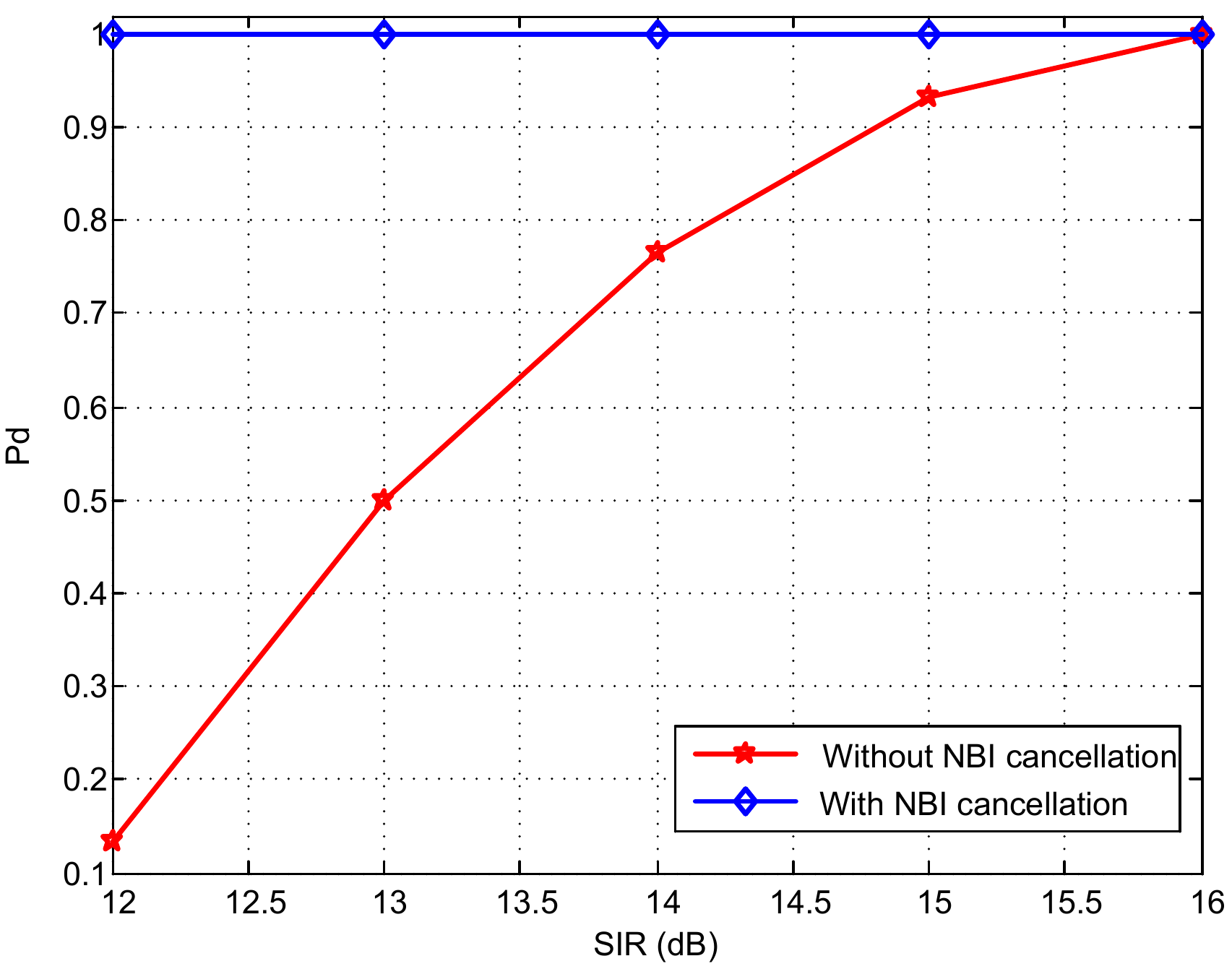}
	\caption{The effect of NBI cancellation}
	\label{cancellation_with_without}
\end{figure}

In order to analyze the performance of the proposed algorithms, several simulations are conducted. For each point, $1 \times {10^4}$ simulations are performed. Fig.~\ref{cancellation_with_without} shows the effect of single tone NBI to compressed sensing system when SNR = 20 dB. Rhombused curve is obtained by~(\ref{system_convex}) while stared curve is obtained by recovering channel information using~(\ref{MIMO_matrix_NBI_convex_opti}). The vertical axis is the probability of detection, which means the probability when correct channel tap location is obtained. The result shows when SIR goes below 15 dB, system performance would be severely influenced by NBI.

\begin{figure}[!t]
	\centering
	\includegraphics[width=2.8in]{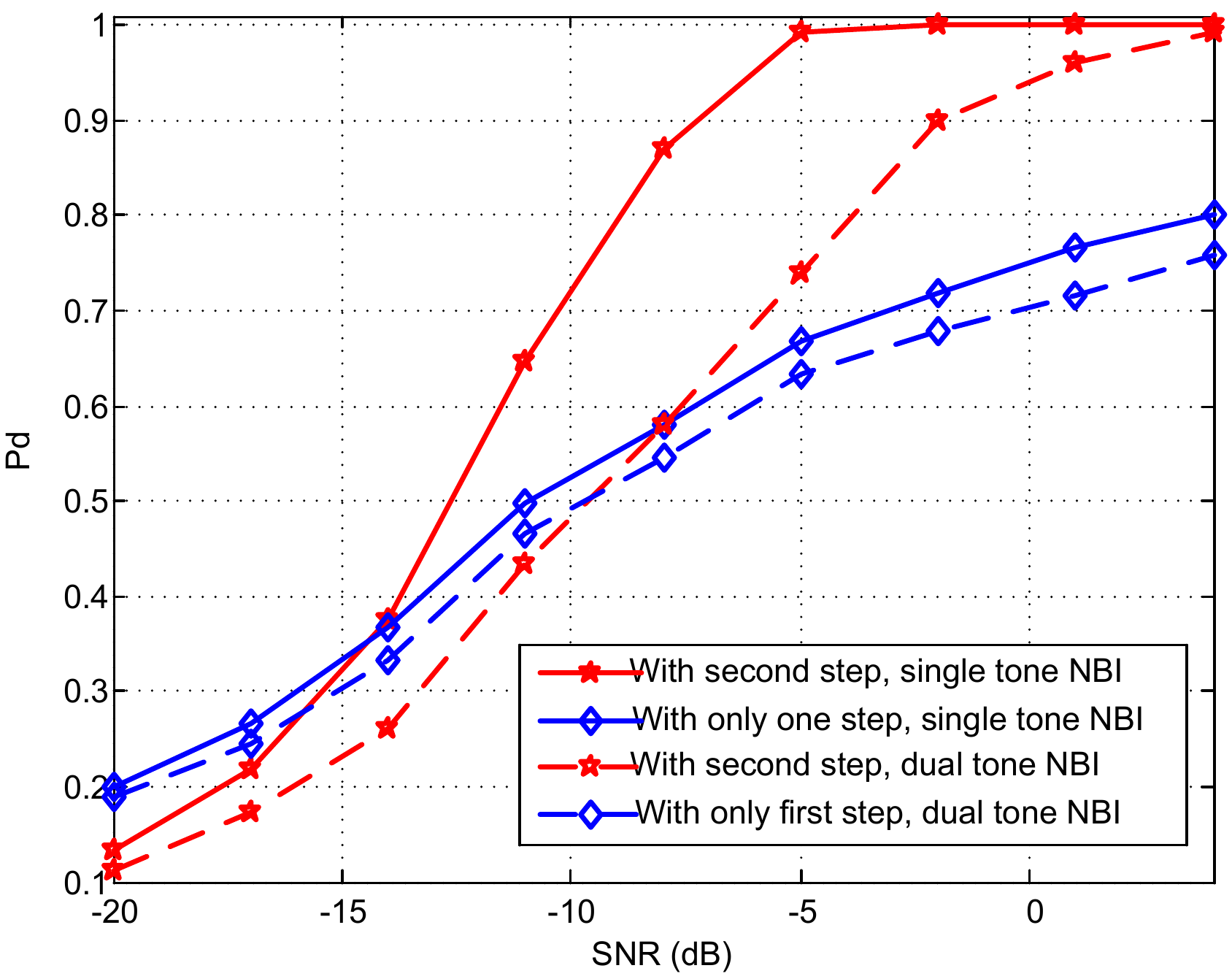}
	\caption{Performance curve of simulation}
	\label{performance}
\end{figure}

With a strong interference (SIR = 0 dB), Fig.~\ref{performance} gives system performance under difference additive white Gaussian noise (AWGN) levels. Three targets situation is considered. Both single tone and dual tone NBIs  results are shown. Two stared curves are for~(\ref{MIMO_matrix_NBI_convex_opti_cancel_convx}). Two rhombused curves are for~(\ref{MIMO_matrix_NBI_convex_opti}). With the second step of channel estimation, the accuracy becomes much higher. In the case of dual-tone, the NBI signal becomes less sparse than the single tone case. Because the CS algorithm highly relies on the sparsity of the signal, the performance is worse than the single tone case.

\section{Conclusion}
\label{conc}


In this letter, we have proposed a CS-based MIMO UWB cognitive radar. An algorithm is designed to get the ToA information from the sub-sampled echoed radar waveform. The algorithm is based on circulant convolution. It can estimate the channel information. Based on the equations of the whole system, a hardware architecture has been proposed to fit into the special structure of the CS system.  NBI signal in the surveillance area is also considered.  An algorithm to estimate and cancel the NBI interference is proposed. The hardware testbed is used to verify the effectiveness of the proposed algorithm in real-world situation. Simulation results show that the proposed algorithms are able to estimate the locations of the jamming interference and provide a performance gain compared to the system without NBI cancellation.

\section*{Acknowledgment}
This work is funded by National Science Foundation
through two grants (ECCS-0901420 and ECCS-0821658), and
Office of Naval Research through two grants (N00010-10-1-
0810 and N00014-11-1-0006). This work is partly funded by an AFOSR subcontract though the prime contractor (RNET Technologies, Inc.) on the CIRE (Center for Innovative Radar Engineering) contract FA8650- 10-D-1750 with AFRL/RY.

\ifCLASSOPTIONcaptionsoff
  \newpage
\fi

\bibliographystyle{IEEEtran}
\bibliography{dsoref/compressed}

\end{document}